\documentclass[12pt]{article}
\usepackage{amssymb}
\usepackage{latexsym}
\usepackage{amsmath}
\usepackage{color}
\usepackage{cite}
\usepackage{units}
\usepackage{tabularx}
\usepackage{graphicx}
\usepackage{cancel}

\numberwithin{equation}{section}

\newcommand{\gsim}{\raisebox{-0.13cm}{~\shortstack{$>$ \\[-0.07cm] $\sim$}}~}

\begin{document}
\pagestyle{empty}
\begin{flushright}
June 2012
\end{flushright}

\begin{center}
{\large\sc {\bf Comment on the Updated CDF ``Ghost'' Events}}  

\vspace{1cm}
{\sc Nicki Bornhauser$^{1}$ and Manuel Drees$^{1}$}  

\vspace*{5mm}
{}$^1${\it Physikalisches Institut and Bethe Center for Theoretical
  Physics, Universit\"at Bonn, Nu\ss{}allee~12, D--53115 Bonn, Germany} 
\end{center}

\vspace*{1cm}
\begin{abstract}
  In 2008 the CDF Collaboration announced the discovery of an excess
  of events with two or more muons, dubbed ``ghost'' events for their
  unusual properties. In a recent update, CDF finds that the azimuthal
  angle distribution between the primary (trigger) muons is
  significantly more back--to--back than that of all known sources of
  di--muon backgrounds. Here we show that this angular distribution
  cannot be reproduced in models where the muons are produced in the
  decays of relatively light $X-$particles: all models of this kind
  also predict a much broader distribution than that found by CDF. We
  conclude that the CDF measurement cannot be described via the
  annihilation of strongly interacting partons, and thus seems to be
  in conflict with basic tenets of QCD.
\end{abstract}

\newpage
\setcounter{page}{1}
\pagestyle{plain}

In 2008 the CDF Collaboration announced the discovery of a very large
sample of events with two or more muons that could not be explained by
known production mechanisms (chiefly heavy flavor and Drell--Yan
production) \cite{bib:ghost}. The two muons with the highest
transverse momenta in an event, the primary muons, have to pass the
following cuts: Each primary muon should have a transverse momentum
$p_T \geq \unit[3]{GeV}$ and a pseudorapidity $|\eta| \leq
0.7$. Additionally their combined invariant mass should be in the
range $\unit[5]{GeV} < m_{\mu\mu} \leq \unit[80]{GeV}$. These
unexplained events were called ``ghost'' events due to their unusual
properties. Among other things, at least one of the two primary
(trigger) muons is produced outside the (inner part of the)
microvertex detector; also, there are roughly equal numbers of
like--sign and opposite--sign primary di--muon pairs. Altogether, CDF
estimated that $84895 \pm 4829$ events with two or more muons belong
to this ``ghost'' category, for an integrated luminosity of
$\unit[742]{pb^{-1}}$. An unexpectedly large fraction of these events
contains additional muons (with $p_T \geq \unit[2]{GeV}$, $|\eta| \leq
1.1$) in $ 36.8^\circ$ cones around the primary muons.

\begin{figure*}
\centering
\resizebox{0.95\textwidth}{!}{
\includegraphics{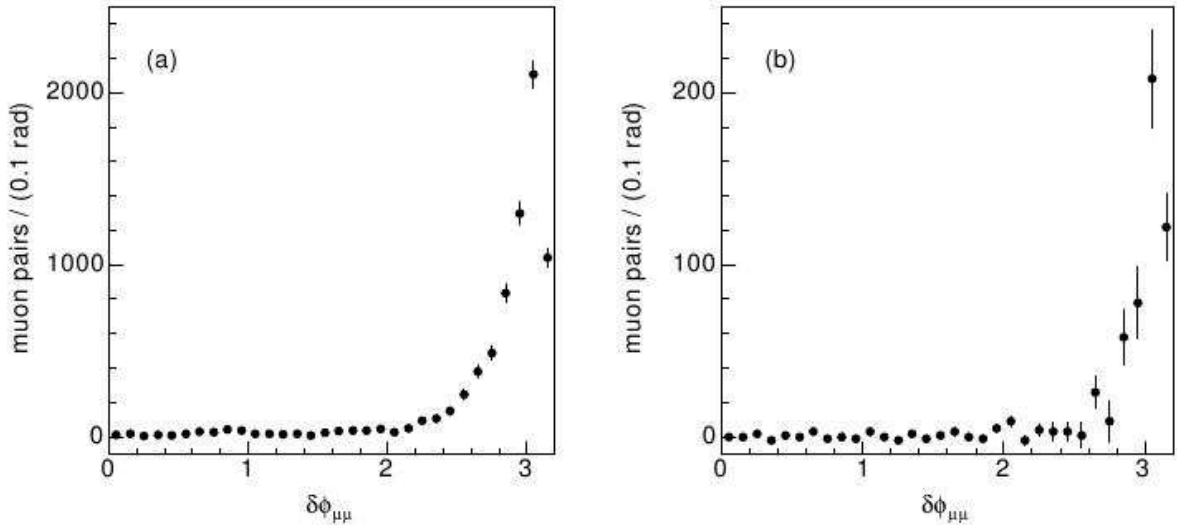}}
\caption{The measured distribution of the azimuthal angle
  $\delta\phi_{\mu\mu}$ between the two primary muons. Each primary
  muon should be accompanied by at least one (\textbf{left}) or two
  (\textbf{right}) additional muons. [Fig. 4 from
  \cite{bib:ghostUpdate}.]}  
\label{fig:deltaPhi_CDF}
\end{figure*}
 
After reviewing their estimate of the QCD production and
investigating a bigger data sample CDF reduced the estimate for the
total number of ghost events to $54437 \pm 14171$ for an integrated
luminosity of $\unit[1426]{pb^{-1}}$ \cite{bib:ghostUpdate}. Note that
this is less than four standard deviations away from zero. However,
$12169 \pm 1319$ of these events have at least one additional
muon. Looking at the azimuthal angle $\delta\phi_{\mu\mu}$ between the
two primary muons they noticed that the distribution differs from
known QCD production for the subset of ghost events where {\em each}
primary muon is accompanied by at least one additional muon, see
Fig.~\ref{fig:deltaPhi_CDF}: most primary muons are very nearly back
to back, whereas conventional sources of multi--muon events have much
broader distributions in $\delta \phi_{\mu\mu}$. CDF concludes from
this that at least this subsample of ``ghost'' events cannot be
explained from known sources, e.g. through mis--identification or late
decays of hadrons.

In 2010 we constructed some models that could explain most of the
(publicly available) features of the original ghost events
\cite{bib:bod3}. We used these models to predict the number
of ``ghost'' events predicted in several older experiments. The fact
that at least one of the muons originated more than $\unit[15]{mm}$
away from the primary vertex, and the large fraction of like--sign
events, should make these events readily recognizable.

In order to simulate ``ghost'' events in Herwig++ \cite{bib:herwig} we
introduced a rather long--lived bosonic $X$ particle, whose pair
production and subsequent decay (with $c \tau_X \gsim \unit[15]{mm}$)
should be responsible for the primary muon pair. $X$ should be a
Majorana particle to account for the equal number of same--sign and
opposite--sign primary muon pairs. We assumed that each $X$ particle
decays into four elementary fermions in order to allow a high number
of muons in the final state. We considered several different
production mechanisms and decay possibilities for the $X$
particles. Within a given scenario the mass and branching ratios of
the $X$ particle were determined by the measured invariant mass
distribution of all muons contained in $36.8^\circ$ cones around the
primary muons and by the measured muon multiplicity distribution.

Here we compute the $\delta\phi_{\mu\mu}$ distribution of the primary
muons for events where each primary muon is accompanied by at least
one additional muon predicted by these models, and compare with the
measurements shown in Fig.~\ref{fig:deltaPhi_CDF}. We first consider a
simple model which reproduces more than $\unit[90]{\%}$ of the
original ghost events with the process $q\bar{q} \rightarrow XX$,
assuming a differential $S-$wave cross section:
\begin{equation*}
\frac{d\sigma(q\bar{q} \rightarrow XX)} {d\cos \theta} =
N_{q\bar{q}} \cdot \frac{\beta}{\hat{s}} = N_{q\bar{q}} \cdot
\frac{ \sqrt{1 - \frac {4 m^2_X} {\hat s} } } {\hat s}. 
\end{equation*}
Here $\hat{s}$ is the squared partonic center of mass energy, $\beta$
is the velocity of the $X$ particles in the partonic center of mass
frame, and $N_{q\bar{q}}$ a constant fitted to the cross section for
``ghost'' events given in ref.\cite{bib:ghost}; the shape of the
$\delta \phi_{\mu\mu}$ distribution does not depend on the value of
this constant. The $X$ mass is $m_X = \unit[1.8]{GeV/c^2}$; see
ref.\cite{bib:bod3} for further details. Our model allows $X$ decays
into one, two our four muons. Since we want to study events where each
primary muon is accompanied by (at least) one additional muon, we here
force both $X$ bosons to decay in two muons with equal probability
either via $X \rightarrow \mu^- \mu^+ u \bar{u}$ or $X \rightarrow
\mu^- \mu^+ d \bar{d}$. We simulated 5 million events, most of which
do not pass the cuts described above.

\begin{figure*}
\centering
\resizebox{0.5\textwidth}{!}{
\includegraphics{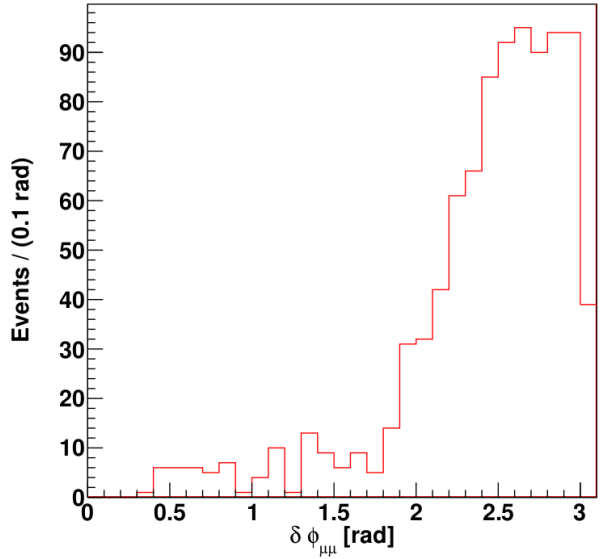}}
\caption{The distribution of the azimuthal angle $\delta\phi_{\mu\mu}$
  between the two primary muons for the simple model. $36.8^\circ$
  cones around the primary muons should contain at least one
  additional muon each.}
\label{fig:deltaPhi_mod1}
\end{figure*}

The predicted $\delta\phi_{\mu\mu}$ distribution after cuts is shown
in Fig.~\ref{fig:deltaPhi_mod1}. Evidently the distribution is much
broader than that shown in the left frame of
Fig.~\ref{fig:deltaPhi_CDF}: this model predicts the cross section to
remain sizable out to $\delta\phi \approx \unit[2]{rad}$, whereas the
CDF distribution essentially cuts off at $\delta\phi \approx
\unit[2.6]{rad}$.

The broadening of the $\delta \phi_{\mu\mu}$ distribution is due to
three physical effects. First, even in the simplest parton level
calculation, only the two $X-$particles are produced back--to--back in
the transverse plane. The decay of the $X$ particles into two (or
more) muons (plus additional particles) generically adds a component
to the $p_T$ vector of the muons which is orthogonal to that of the
parent $X-$particle. Second, perturbative gluon radiation, in
particular initial state radiation, gives a transverse ``kick'' to the
$X$ pair. However, this is not sufficient to describe, e.g.,
Drell--Yan production adequately in the region of small $p_T$ of the
lepton pair. To this end, event generator programs allow the colliding
partons to have an ``intrinsic'' $p_T$ of typically one or two GeV,
which gives an additional transverse kick to the $X$ pair; we modeled
this using the default settings of Herwig++. The second and third
effect therefore lead to a transverse opening angle between the two
$X$ particles that is less than $\pi$; for given $p_T$ of the $X$ pair
this effect will be smaller for larger $p_T$ of each $X$ particle.

\begin{figure*}
\centering
\resizebox{0.5\textwidth}{!}{
\includegraphics{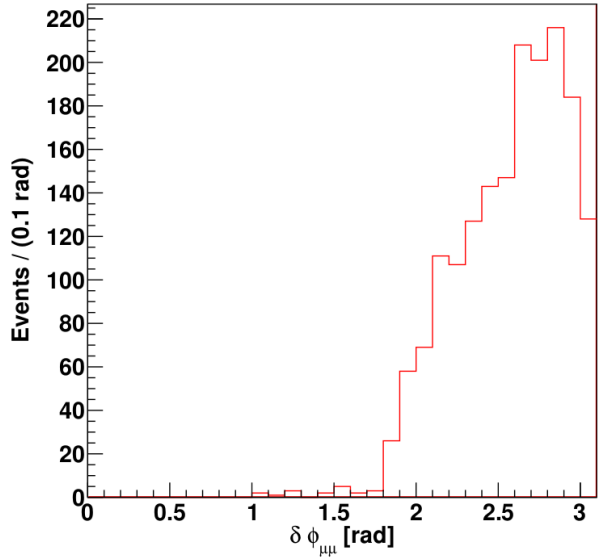}}
\caption{As in Fig.~2, but for the more sophisticated model.}  
\label{fig:deltaPhi_mod2}
\end{figure*}

This simple model does not describe the original ghost events with
higher muon multiplicities very well. This includes the invariant mass
distribution of all muons in events where each primary muon is
accompanied by at least one additional muon: the distribution
predicted by this simple model peaks at smaller values, and drops off
faster towards large values, than the distributions published in the
original ``ghost'' analysis do \cite{bib:bod3}. In order to produce
more events with large invariant mass, we therefore also consider a
more sophisticated model, where $X$ pair production proceeds via a
broad resonance $Y$. The corresponding cross section is
\cite{bib:bod3}
\begin{equation*}
\frac{d\sigma(q\bar{q} \rightarrow Y \rightarrow XX)} {d\cos
  \theta} = N_{q\bar{q}}^{BW} \cdot \frac{\hat{s}^2}{(\hat{s} -
  m_Y^2)^2 + \Gamma_Y^2 m_Y^2} \cdot \frac{\sqrt{1 - \frac{4
      m^2_X}{\hat{s}}}}{\hat{s}},
\end{equation*}
where the constant $N_{q\bar{q}}^{BW}$ is again chosen to match the
original ghost cross section. $Y$ is a resonance with the mass $m_Y =
\unit[110]{GeV}$ and width $\Gamma_Y = \unit[110]{GeV}$, and $m_X =
\unit[4.6]{GeV}$. In ref.\cite{bib:bod3} we also modified the $X$
decay branching ratios relative to the simple model, allowing decays
where muon number is violated; this allowed us to reproduce the large
fraction of ``ghost'' events where a primary muon is accompanied by a
nearby secondary muon with the same charge. However, for the purpose
of the present analysis the charges of the muons are irrelevant, so we
focus on the decay mode $X \rightarrow \tau^+ \tau^- \mu^+ \mu^-$. The
simulation of 5 million events leads to the $\delta\phi_{\mu\mu}$
distribution after cuts shown in Fig.~\ref{fig:deltaPhi_mod2}.

Evidently this model does not describe the measured distribution shown
in Fig.~\ref{fig:deltaPhi_CDF}, either; the distribution is again much
broader than the measurement, although slightly narrower than that
predicted by the simple model. Note that in spite of the larger value
of $m_X$ the $Q-$value of the decay is actually somewhat smaller than
in the simple model, reducing the difference between $\delta
\phi_{XX}$ and $\delta \phi_{\mu\mu}$. Moreover, due to the
Breit--Wigner form of the cross section, $X-$particles are now often
produced with significant transverse momentum. This is required to
reproduce the multi--muon invariant mass distribution shown in
ref.\cite{bib:ghost}; it also reduces the effect of a fixed transverse
``kick'', e.g. due to initial state radiation, on $\delta
\phi_{\mu\mu}$.

It seems very difficult to generate a significantly narrower $\delta
\phi_{\mu\mu}$ distribution, as required to reproduce the
distributions shown in \cite{bib:ghostUpdate}, without distorting the
multi--muon invariant mass distributions within the cones around the
primary muons or within the entire event, and/or without violating
basic tenets of QCD. In particular, further increasing the average
$p_T$ of the $X-$particles would increase the invariant mass of all
muons in the event too much; conversely, further decreasing the
$Q-$value of $X-$decays would make the invariant mass distribution of
muons inside the cones around the primary muons too
soft.\footnote{Here we are assuming that the improved background
  estimate of ref.\cite{bib:ghostUpdate} does not greatly distort the
  shapes of these invariant mass distributions presented in the
  original analysis \cite{bib:ghost}. Unfortunately the update
  \cite{bib:ghostUpdate} only presents distributions in $\delta
  \phi_{\mu\mu}$; no invariant mass distributions are given.} Finally,
choosing a $q \bar q$ initial state already minimizes QCD initial
state radiation; gluons radiate considerably more, leading to an even
broader distribution.

We therefore conclude that it is very difficult, if not impossible, to
reproduce the $\delta \phi_{\mu\mu}$ distribution of the CDF ``ghost''
events within a quantum field theoretical model.

\subsection*{Acknowledgments}
NB wants to thank the ``Bonn-Cologne Graduate School of Physics and Astronomy''
and the ``Universit\"at Bonn'' for the financial support. This work
was partially supported by the BMBF--Theorieverbund.


\begin{thebibliography}{500}
\bibitem{bib:ghost}
T. Aaltonen \textit{et. al.} (CDF Collaboration),
arXiv:0810.5357v2 [hep-ex]

\bibitem{bib:ghostUpdate}
T. Aaltonen \textit{et. al.} (CDF Collaboration),
arXiv:1111.5242v1 [hep-ex]

\bibitem{bib:bod3}
N. Bornhauser and M. Drees, {\em Eur. Phys. J.} \textbf{C71} (2011) 1581
[arXiv:1009.4749v1 [hep-ph]].

\bibitem{bib:herwig}
M. B\"ahr \textit{et. al.}, {\em Eur. Phys. J.} \textbf{C58} (2008) 639
[arXiv:0803.0883 [hep-ph]].
\end{thebibliography}
\end{document}